\documentclass[a4paper]{article}

\usepackage{icrc2013}
\usepackage[english]{babel}
\usepackage{amsmath, amssymb}
\usepackage{epsf}

\title{MAGIC Gamma-ray Observations of the Perseus Galaxy Cluster}

\shorttitle{MAGIC Gamma-ray Observations of the Perseus Galaxy Cluster}

\authors{
Fabio Zandanel$^{1,2}$,
Pierre Colin$^{3}$,
Saverio Lombardi$^{4}$,
Michele Doro$^{5}$,
Dorit Eisenacher$^{6}$,
Dorothee Hildebrand$^{7}$,
Francisco Prada$^{1,8,9}$
for the MAGIC Collaboration; and
Christoph Pfrommer$^{10}$,
Anders Pinzke$^{11}$
}

\afiliations{
$^1$ Instituto de Astrof\'isica de Andaluc\'ia (IAA-CSIC), E-18080 Granada, Spain\\
$^2$ now at: GRAPPA Institute, University of Amsterdam, N-1098XH Amsterdam, Netherlands\\
$^3$ Max-Planck-Institut f{\"u}ur Physik, D-80805 M{\"u}nchen, Germany\\
$^4$ INAF National Institute for Astrophysics, I-00136 Rome, Italy\\
$^5$ Universit\`a di Padova and INFN, I-35131 Padova, Italy\\
$^6$ Universit{\"a}t W{\"u}rzburg, 97074 W{\"u}rzburg, Germany\\
$^7$ ETH Zurich, 8093 Zurich, Switzerland\\
$^8$ Campus of International Excellence UAM+CSIC, Cantoblanco, E-28049 Madrid, Spain\\
$^9$ Instituto de F\'sica Te\'orica, (UAM/CSIC), Universidad Aut\'onoma de Madrid, Cantoblanco, E-28049 Madrid, Spain\\
$^{10}$ Heidelberg Institute for Theoretical Studies (HITS), D-69118 Heidelberg, Germany\\
$^{11}$ University of California - Santa Barbara, Department of Physics, CA 93106-9530, USA\\
}

\email{f.zandanel@uva.nl}

\abstract{In order to detect the gamma-ray emission from cosmic ray (CR) interactions with the intra-cluster medium, the ground-based imaging Cherenkov telescope MAGIC conducted the deepest-to-date observational campaign targeting a galaxy cluster at very high-energies ($\gtrsim100$~GeV) and observed the Perseus cluster for a total of 85~hr during $2009$--$2011$. The observations constrain the average CR-to-thermal pressure ratio to be \mbox{$1$--$2$\%} and the maximum CR acceleration efficiency at structure formation shocks to be $<50$\%. Alternatively, this may argue for non-negligible CR transport processes such as CR streaming and diffusion into the outer cluster regions. Additionally, assuming that the Perseus radio mini-halo is generated by secondaries created in hadronic CR interactions, the central magnetic field is limited to be $>4$--$9$~$\mu$G. This range is well below the field strength inferred from Faraday rotation measurements and, therefore, the hadronic model remains a plausible explanation of the Perseus radio mini-halo. Following this successful campaign, MAGIC is continuing collecting data on Perseus.}


\begin{document}
\maketitle

\section{Introduction}

In the standard hierarchical model of structure formation, clusters of galaxies are
the largest and latest objects to form through mergers of smaller galaxy groups. 
During these merger events, enormous amounts of energy --- of the order of the final gas binding 
energy $E_\mathrm{bind} \sim 3 \times (10^{61}$--$10^{63})$~erg --- are released through
turbulence and collisionless shocks. This energy is dissipated on a dynamical 
timescale of about  $1$~Gyr and the corresponding energy dissipation 
rates are $L \sim (10^{45}$--$10^{47})~\mbox{erg~s}^{-1}$ (see \cite{bib:1} for a review). 
If only a small fraction of this energy is conveyed into non-thermal particles, 
the associated emission should be detectable in gamma-rays and could be 
used to decipher the history of structure formation.

Many galaxy clusters show large scale diffuse synchrotron radio emission in the
form of so-called radio \mbox{(mini-)halos} which proves the existence of magnetic
fields and relativistic electrons permeating the intra-cluster medium (ICM)
\cite{bib:2}. Galaxy clusters should also be acceleration sites for relativistic protons 
and heavier relativistic nuclei, similarly to shocks within our Galaxy such as those 
in supernova remnants. Protons and heavier nuclei are accelerated more efficiently 
to relativistic energies with respect to electrons because of their higher masses. Therefore, we 
expected a ratio of the spectral energy flux of cosmic ray (CR) protons to CR electrons above 
1~GeV of about 100 as it is observed in our Galaxy between $1$--$10$~GeV \cite{bib:3}. 
CR protons also have cooling times that are longer than the age of the Universe, and hence 
can accumulate over the Hubble time in a galaxy cluster \cite{bib:4}. For typical gas density 
$n\sim10^{-3}$~cm$^{-3}$ in galaxy clusters, the radiative cooling time of CR protons 
is much longer than the hadronic timescale, 
$\tau_\mathrm{pp}\simeq 30 \,\mathrm{Gyr}\times (n/10^{-3}\,\mathrm{cm}^{-3})^{-1}$, 
on which CR protons collide inelastically with ambient gas protons.
This is the process in which we are primarily interested here as it generates charged and 
neutral pions that successively decay into synchrotron-emitting electrons/positrons and 
gamma-rays, respectively.

The Perseus cluster was selected for the Major Atmospheric Gamma Imaging Cherenkov (MAGIC) 
observations as it is the most promising target for the detection of gamma-rays resulting from neutral 
pions produced in hadronic CR interactions with the ICM \cite{bib:5,bib:6,bib:7,bib:8}. This cluster 
of galaxies, at a distance of 77.7~Mpc (z = 0.018), is the brightest X-ray cluster with a luminosity in the 
soft X-ray band of $L_{0.1-2.4~\mathrm{keV}} = 8.3\times10^{44}$~erg~s$^{-1}$ \cite{bib:9}. 
It contains a massive cool core with high central gas densities of about $0.05$~cm$^{-3}$ \cite{bib:10} and 
a luminous radio mini-halo with an extension of 200~kpc \cite{bib:11}.

\newpage

\section{MAGIC Observations and Data Analysis}

The MAGIC telescopes are two 17\,m dish IACTs located at the Roque de los
Muchachos observatory ($28.8^\circ$N, $17.8^\circ$W, 2200~m a.s.l.), on the
Canary Island of La Palma. At the energies of interest here, i.e, above approximately 
$600$~GeV, the sensitivity in 50~hr of observations is $\sim 0.7\%$ of the Crab Nebula 
flux and the point spread function (PSF), defined as a 2-dimensional Gaussian, has 
a $\sigma\simeq0.06^\circ$ \cite{bib:12}.

The Perseus cluster region was observed by the MAGIC telescopes from
October 2009 to February 2011 for a total of about $99$~hr \cite{bib:8,bib:13,bib:14}. 
Observations were performed in wobble mode, tracking positions $0.4^\circ$ from the 
cluster center at low zenith angles ($12^\circ$--$36^\circ$).
The data quality check resulted in the rejection of about $14.4$~hr of data,  
mainly due to non-optimal atmospheric conditions, and the final data sample 
consists of $84.5$~hr of effective observation time. The standard MAGIC
stereo analysis chain was used for calibration and image cleaning \cite{bib:12}. 

\begin{figure}[t]
\centering
\includegraphics[width=0.5\textwidth]{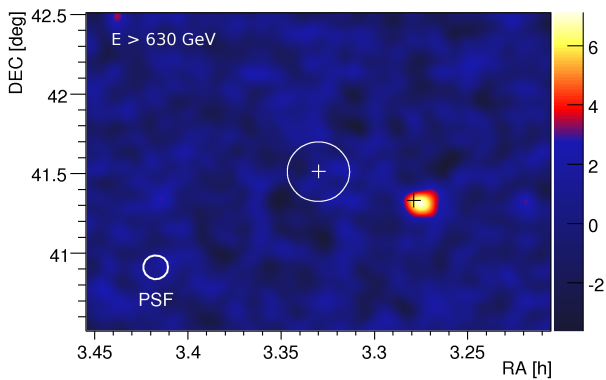}
\caption{
Significance skymap of the Perseus cluster region above 630~GeV. The white cross 
marks the NGC~1275 position, while the black cross marks the IC~310 one.
The central white circle shows the region of 0.15$^\circ$ radius used to derive 
constraints on the CR-induced gamma-ray emission.
}
\label{skymap}
\end{figure}

Cosmological simulations suggest that the spectrum of CR-induced gamma-rays 
obeys a power-law, $F\propto E^{-\alpha}$, with a spectral index of about $\alpha=2.2$ 
at the energies of interest here \cite{bib:5,bib:6}. The corresponding signal is extended. 
However, because of the dense gas in the cluster center, approximately 60$\%$ of the 
emission is coming from a region centered on NGC~1275 with a radius of $0.15^\circ$. 
NGC~1275 is a radio galaxy located at the center of Perseus and its emission is dominant below 
approximately 600~GeV with a spectral index of about 4 \cite{bib:14}. 
Therefore, we limit the analysis to energies above $630$~GeV where the NGC~1275 
signal is not detected. Figure~\ref{skymap} shows the significance skymap above 630~GeV. 
In contrast to NGC~1275, the spectrum of IC~310, another radio galaxy in the cluster, is very 
hard and remains detectable above 600~GeV \cite{bib:13}. IC~310 is $\sim 0.6^\circ(\simeq 10$~PSF) 
away from the cluster center and its highly variable emission does not leak into the signal region. Nevertheless, it 
can affect the background estimation and, therefore, we measure it with three off-source positions 
at 0.4$^\circ$ from the the camera center and $>0.28^\circ$ away from IC~310 guaranteeing that 
there is no contamination.

\begin{figure}[t]
\centering
\includegraphics[width=0.48\textwidth]{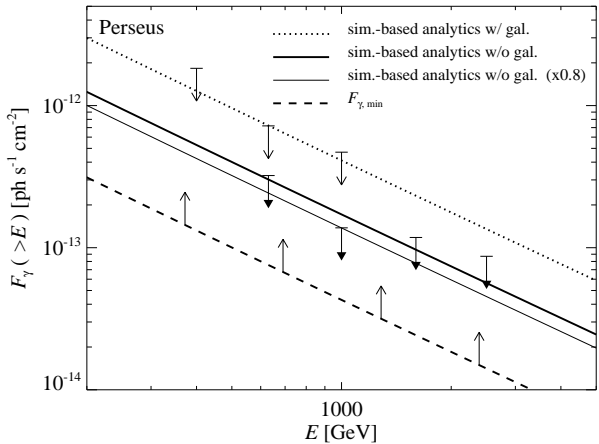}
\caption{
The solid arrows show the integral flux ULs obtained in this work \cite{bib:8}, while 
the upper arrows show those obtained from single telescope observations 
(point-like ULs; \cite{bib:5}). We compare ULs with the simulated spectra of
the CR-induced gamma-ray emission in the Perseus cluster coming from within 
a radius of $0.15^\circ$ around the center. We show both the conservative model 
without galaxies (solid line) and the model with galaxies (dotted line). The flux UL 
at 1~TeV is a factor of 1.25 below the conservative model and, therefore, constrains 
a combination of CR acceleration efficiency and transport processes. We also show 
the minimum gamma-ray flux estimates for the hadronic model of the Perseus 
radio mini-halo (dashed line with minimum flux arrows) adopting the 
simulation-motivated spectral index of  2.2.
}
\label{spectrum}
\end{figure}

As clear from Figure~\ref{skymap}, we do not detect any gamma-ray emission
coming from the center of the Perseus cluster above 630~GeV. Therefore, we derive integral flux upper 
limits (ULs) for several energy thresholds and for a spectral index of $2.2$; we 
show them in Table~\ref{ULs}. The ULs have been corrected to take into account the 
expected source extension comparing the fraction of the total events inside the signal 
region for a point-like source and for the expected CR-induced signal. Therefore, the ULs 
shown in Figure~\ref{spectrum} can be compared with the theoretical expectations for the 
region within a radius of $0.15^\circ$. The UL estimation is performed using the Rolke
method \cite{bib:15} with a confidence level of $95$\% and a total systematic uncertainty of $30$\%. 
The integral UL for energies above 1~TeV corresponds to the best sensitivity for sources 
with spectral index $2.2$ and it is the most constraining value. Therefore, we will adopt 
the $1$~TeV UL as reference value for the following discussion. 

\begin{table}[t]
\begin{center}
\begin{tabular}{ccccc}
\hline
\phantom{\Big|}
E$_{\mathrm{th}} [\mathrm{GeV}]$ & $\sigma_{\mathrm{LiMa}}^{\mathrm{PL}}$ & N$_{\mathrm{UL}}^{\mathrm{PL}}$ & F$_{\mathrm{UL}}^{\mathrm{PL}}$ & F$_{\mathrm{UL}}^{0.15^{\circ}}$\\
\hline
630    &   0.59  &  84.7  &   2.93  &  3.22\\
1000  &   0.15  &  41.4  &  1.25  & 1.38\\
1600  &   0.33  &  38.7  &  1.07  &  1.18\\ 
2500  &   0.38  &  28.8  &  0.79  &  0.87 \\  
\hline
\end{tabular}
\caption{Integral flux ULs F$_{\mathrm{UL}}$ for a power-law gamma-ray 
spectrum with spectral index $2.2$, above a given energy threshold
E$_{\mathrm{th}}$, in units of $10^{-13}$~cm$^{-2}$~s$^{-1}$. We 
show both the point-like (PL) and the $0.15^\circ$ region cases. We also 
show the corresponding significance $\sigma_{\mathrm{LiMa}}^{\mathrm{PL}}$ 
and ULs in number of events N$_{\mathrm{UL}}^{\mathrm{PL}}$ (before 
applying the source extension correction).}
\label{ULs}
\end{center}
\end{table}

\section{Results and Discussion}

In order to model the thermal pressure of the cluster, we adopt the measured electron 
temperature and density profiles of Perseus \cite{bib:10}. 

Considering a simplified analytical model that assumes a power-law CR momentum spectrum 
and a constant CR-to-thermal pressure ratio \cite{bib:16}, we constrain this last quantity, 
$X_\mathrm{CR} = P_{\mathrm{CR}}/P_\mathrm{th}$ (averaged within the virial radius of 2~Mpc), 
to be $<0.77$\% and 11.6\% for $\alpha$ varying between 2.1 and 2.5.  For a spectral index 
of 2.2, favored by simulations, we obtain $X_\mathrm{CR}<1.1\%$.

For a more realistic approach, we turn to cosmological hydrodynamical simulations and
adopt the semi-analytical model developed by Pinzke \& Pfrommer \cite{bib:6}. The 
normalization of the gamma-ray emission scales with the CR maximum acceleration efficiency 
at shocks. Motivated by recent observations and theoretical studies of supernova remnants, 
we assume that 50\% of the dissipated energy at strong shocks is injected into CRs, while 
this efficiency rapidly decreases for weaker shocks. 

These cosmological simulations only consider advective transport of CRs by turbulent gas 
motions which produces a centrally enhanced profile. However, CR transport phenomena, such 
as CR diffusion and streaming, flattens the CR radial profile, producing a spatially constant CR 
number density in the most extreme case \cite{bib:17,bib:18,bib:19}. This results in a 
bimodality of the CR spatial distribution, with merging (relaxed) clusters showing a centrally 
peaked (flat) CR density. As a consequence, relaxed clusters could have a reduced gamma-ray 
luminosity with respect to the simulation predictions.

In order to assess the biases associated with the insufficient numerical resolution of the simulations, 
as well as incompletely understood physical properties of the cluster plasma, we performed our analysis 
with two limiting cases bracketing the realistic case (see \cite{bib:5,bib:8} for details).
In our optimistic CR model (simulation-based analytics with galaxies), we calculated the cluster total 
gamma-ray flux within a given solid angle, while we cut the emission from individual galaxies 
and compact galactic-sized objects in our more conservative model (simulation-based analytics without 
galaxies). Figure~\ref{spectrum} shows the corresponding spectrum for Perseus 
within an aperture of radius 0.15$^\circ$. The MAGIC UL above 1~TeV falls below the flux level 
of the conservative model by 20\%. Therefore, we find that $X_\mathrm{CR}$ of the simulation-based 
analytical model have to to be less than 1.6\% within 0.15$^\circ$ (200~kpc). Assuming this spatial CR 
profile, we obtain a CR-to-thermal pressure ratio $< 1.7\%$ within the virial radius and $< 5\%$ 
within 20~kpc. 

For the first time, the CR physics of galaxy clusters in simulations can be constrained.
The MAGIC results either limit the maximum acceleration efficiency of CRs at strong structure formation shocks 
to $<50\%$ or indeed indicate possible CR streaming and diffusion out of the cluster core region. 

For clusters hosting diffuse radio emission, we can estimate the minimum gamma-ray flux in
the hadronic scenario for radio halos. This suggests that the radio mini-halo of Perseus
\cite{bib:11} is produced by secondary electrons coming from the CR hadronic interactions
with the ICM. Assuming that this is the case, and considering that a stationary distribution 
of CR electrons loses all its energy to synchrotron radiation for strong magnetic fields 
$B \gg B_\mathrm{CMB}$ \cite{bib:8}, we can derive a minimum gamma-ray flux. 
This results to be a factor of 1.8 to 17.3 lower than our ULs for a spectral index between 
$2.1 \leq \alpha \leq 2.5$. The case of $\alpha=2.2$, as suggested by simulations,
is shown in Figure~\ref{spectrum}. This is within the reach of future observations
as the current MAGIC ULs are only a factor 3.2 higher.

We can turn this last argument around and use our gamma-ray ULs to derive a lower 
limit on the magnetic field value needed to explain the observed synchrotron radio 
emission within the hadronic scenario. The lower the gamma-ray limit, the higher 
the magnetic field needed to generate the radio emission. 

First, we match the radio emission profile fixing the radial CR distribution for a given
magnetic field model, $B(r) = B_{0} \,(n(r)/n(0))^{\alpha_B}$.
We then require the gamma-ray flux from pion decays to match the MAGIC ULs, fixing the 
normalization of the CR distribution, and eventually obtain the lower limits on the magnetic 
field central value $B_{0, min}$ shown in Table~\ref{magnetic}. They are in 
the 2--13~$\mu$G range for the values of $\alpha$ and $\alpha_B$ used in this study 
and suggested by radio observations. Both the thermal equipartition value in the center 
of Perseus, $B_\mathrm{eq,0}\simeq 80\,\mu$G, and Faraday rotation measurement (RM) 
estimates \cite{bib:20} are much higher than the magnetic field values obtained here. 
We therefore conclude that there is still considerable available phase space for the 
hadronic model as an explanation of the Perseus radio mini-halo emission.

\begin{table}[t]
\begin{center}
\begin{tabular}{ccccc}
\hline
\phantom{\Big|}
           & \multicolumn{4}{c}{Minimum magnetic field $B_{0, min}$~$[\mu\mathrm{G}]$} \\
$\alpha_B$ & \multicolumn{4}{c}{$\alpha$}\\
           & \quad 2.1 \quad & \quad 2.2 \quad & \quad 2.3 \quad & \quad 2.5 \quad \\
\hline\\[-0.5em]
0.3 & 5.86 & 4.09 & 3.15 & 2.06 \\ 
0.5 & 8.62 & 6.02 & 4.63 & 3.05 \\
0.7 & 13.1 & 9.16 & 7.08 & 4.68 \\[0.25em]
\hline
\end{tabular}
\caption{Constraints on the central magnetic field value in the hadronic scenario for radio (mini-)halos.}
\label{magnetic}
\end{center}
\end{table}

\section{Conclusions}

MAGIC observed the Perseus cluster -- the best target where to search for
pion decays from CR hadronic interactions with the ICM -- for a total of $85$~hr 
between October 2009 and February 2011 \cite{bib:8}. This campaign represents the longest 
observation ever of a galaxy cluster at very high-energies ($\gtrsim100$~GeV) and resulted 
in the detection of the IC~310 \cite{bib:13,bib:21} and NGC~1275 \cite{bib:14,bib:22} radio
galaxies. No significant excess of gamma-rays was detected from the cluster central region 
at energies above 630~GeV where the photon yield from NGC~1275 is negligible.

Using a simplified analytical approach, that assumes a power-law CR momentum 
spectrum and a constant CR-to-thermal pressure ratio, we can constrain 
$X_\mathrm{CR}< 0.8\%$ and 12\% for $\alpha$ varying between 2.1 and 2.5. 
For a spectral index of $\alpha=2.2$, favored by simulations, we find that $X_\mathrm{CR}<1.1\%$. 

Adopting the simulation-based approach \cite{bib:6}, we obtain $X_\mathrm{CR}<1.7\%$. 
This is a factor of 1.25 below the model and -- for the first time -- limits the underlying 
physics of the simulation. This could either indicate that the CR maximum acceleration efficiency 
at strong structure formation shocks is $<50\%$ or may point to CR streaming and diffusion out 
of the cluster core region. This would lower the central $X_\mathrm{CR}$ values and, correspondingly,
the gamma-ray emission.

Adopting a strong magnetic field, $B\gg B_\mathrm{CMB}$, everywhere in the radio-emitting 
region, we can estimate the minimum gamma-ray flux in the hadronic model of radio 
\mbox{(mini-)halos}. This is a factor of 2 to 18 below the MAGIC ULs for $\alpha$ varying 
between 2.1 and 2.5. For $\alpha=2.2$, as suggested by simulations, the minimum 
gamma-ray flux is a factor of 3.2 lower than the MAGIC ULs, within the reach of future observations.  

Finally, by matching the radio emission profile and requiring the pion-decay 
gamma-ray flux to match the MAGIC ULs, we obtain lower limits on the central
magnetic field value in Perseus. The inferred values are $2~\mu\mathrm{G} \leq
B_{0,\mathrm{min}} \leq13~\mu$G for the explored parameter space. Since this
is smaller than recent field strengths estimates through Faraday RM studies 
in cool core clusters, the hadronic model is an interesting possibility 
for explaining the radio mini-halo emission. 

This shows the potential of future gamma-ray observations of the Perseus cluster
of galaxies to refine the parameters of the hadronic model, assessing its validity in 
explaining radio (mini-)halos. Indeed, the detection of the CR-induced gamma-ray
emission could be within the reach of future observations. For this reason, MAGIC is
extending this successful observation campaign and already accumulated additional
$100$~hr during $2012$--$2013$.

\vspace*{0.5cm}
\footnotesize{{\bf Acknowledgment:}{
We would like to thank the Instituto de Astrof\'{\i}sica de
Canarias for the excellent working conditions at the
Observatorio del Roque de los Muchachos in La Palma.
The support of the German BMBF and MPG, the Italian INFN, 
the Swiss National Fund SNF, and the Spanish MICINN is 
gratefully acknowledged. This work was also supported by the CPAN CSD2007-00042 and MultiDark
CSD2009-00064 projects of the Spanish Consolider-Ingenio 2010
programme, by grant 127740 of  the Academy of Finland,
by the DFG Cluster of Excellence ``Origin and Structure of the 
Universe'', by the DFG Collaborative Research Centers SFB823/C4 and SFB876/C3,
and by the Polish MNiSzW grant 745/N-HESS-MAGIC/2010/0.}}


\end{document}